\documentclass[twocolumn,prb,preprintnumbers,amsmath,amssymb,floatfix]{revtex4-1}
\usepackage{graphicx}
\usepackage{amsmath}
\usepackage{amssymb}
\usepackage{color}
\usepackage{cases}

\begin{document}

\title{Tree-ansatz percolation of hard spheres}

\author{Claudio Grimaldi}\email{claudio.grimaldi@epfl.ch}
\affiliation{Laboratory of Physics of Complex Matter, Ecole Polytechnique F\'ed\'erale
de Lausanne, Station 3, CP-1015 Lausanne, Switzerland}

\begin{abstract}
Suspensions of hard core spherical particles of diameter $D$ with inter-core connectivity range $\delta$
can be described in terms of random geometric graphs, where nodes represent the
sphere centers and edges are assigned to any two particles separated by a distance smaller
than $\delta$. By exploiting the property that closed loops of connected spheres becomes
increasingly rare as the connectivity range diminishes, we study continuum percolation of hard spheres 
by treating the network of connected particles as having a treelike structure for small $\delta/D$. 
We derive an analytic expression of the percolation threshold which becomes increasingly accurate as $\delta/D$ diminishes,
and whose validity can be extended to a broader range of connectivity distances by a simple rescaling.
\end{abstract}

\maketitle

\section{Introduction}
\label{sec:intro}

In random multi-phase heterogeneous systems, in which objects or particles may be connected according to some connectedness criterion,
the percolation threshold marks the occurrence of a macroscopic, system-spanning cluster or component of connected objects.\cite{Stauffer1994,Torquato2002}
In most systems of practical interest, percolation is  
established among particles that occupy random positions in space and that are not constrained by an underlying regular 
lattice. In such continuum percolating systems, the critical threshold depends on the shape, orientation, and size distributions
of the percolating objects as well as on the connectedness criterion. 

While the influence of these factors on the critical threshold is mostly studied by means 
of numerical simulations of finite systems, and analytical approximate results exist for many models of heterogeneous 
systems,\cite{Torquato2002} exact analytical expressions for the percolation threshold have been found only for a few continuum 
percolation systems in some asymptotic limit.
Notable examples are dispersions of isotropically oriented penetrable rods (or cylinders, spherocylinders) with asymptotically large aspect 
ratios,\cite{Bug1985} and penetrable hyperspheres and oriented hypercubes in the infinite dimensional limit.\cite{Penrose1996,Torquato2012} 
Exact results on the percolation threshold of these families of models have been found also for generalized case of particle size 
polydispersity\cite{Otten2009,Grimaldi2015a,Chatterjee2015} and, for the case of rod dispersion, for hard-core impenetrability.\cite{Otten2009} 

The  property that crucially makes these classes of models exactly tractable is the statistical irrelevance of the contribution of 
closed loops of connected particles to the incipient percolating cluster. This property allows the connectivity network
to have a dendritic, treelike structure, which allows for a closed form solution for the percolation threshold.

In this article it is shown that the percolating network of hard spherical particles 
displays a similar property. Namely, the probability of finding closed loops becomes smaller as the connectivity distance between the 
hard spheres is reduced. This observation enables us to derive a theory of percolation of hard spherical particles in which the 
connectivity network is treated as having a treelike structure, while at the same time the many-body correlations 
of the hard-sphere fluid are fully preserved. The resulting percolation transition agrees well with existing numerical results in the 
limit of short connectivity distance, and an analytic expression of the critical threshold is derived which has a much broader 
range of validity.

\section{Clustering coefficient for hard spheres}
\label{clustering}

We start by considering a random dispersion of $N$ hard-sphere particles of diameter $D$ which occupy a three dimensional region of volume $V$.
The fractional volume occupied by the spheres is $\phi=\pi D^3\rho/6$, where $\rho=N/V$ is the number density.
A pair of spheres centered at $\mathbf{r}_i$ and $\mathbf{r}_j$ are considered as connected to each other
if the distance between their center, $r_{ij}=\vert\mathbf{r}_i-\mathbf{r}_j\vert$, is smaller than $\Delta=D+\delta$, 
where $\delta\geq 0$ is a given distance between the closest surfaces of the spheres.  
This connectivity criterion can be equivalently formulated by associating to each impenetrable
sphere a concentric penetrable shell of thickness $\delta/2$. In the resulting system of composite particles, often referred to
as the cherry-pit model,\cite{Torquato2002,Torquato1984} a pair of semi-penetrable spheres are therefore connected if their penetrable
shells overlap. 

Next, we construct a random geometric graph, or network, whose nodes (or vertices) are associated to the 
centers of the spheres, and edges (or links) between pairs of nodes are assigned according to the aforementioned 
connectivity criterion. The probability that two nodes are directly linked by an edge is therefore given by
\begin{equation}
\label{prob1}
p=\frac{1}{N(N-1)}\left\langle\sum_{i,j}'\theta(\Delta-r_{ij})\right\rangle,
\end{equation}
where $\left\langle\cdots\right\rangle$ denotes the ensemble average over the configurations of the $N$ hard-sphere system,
$\theta(x)=1$ for $x\geq 0$ and $\theta(x)=0$ for $x<0$ is the Heaviside step function, and the prime symbol over the summation 
indicates that the terms with equal indexes must be omitted.

\begin{figure}[t]
\begin{center}
\includegraphics[scale=0.38,clip=true]{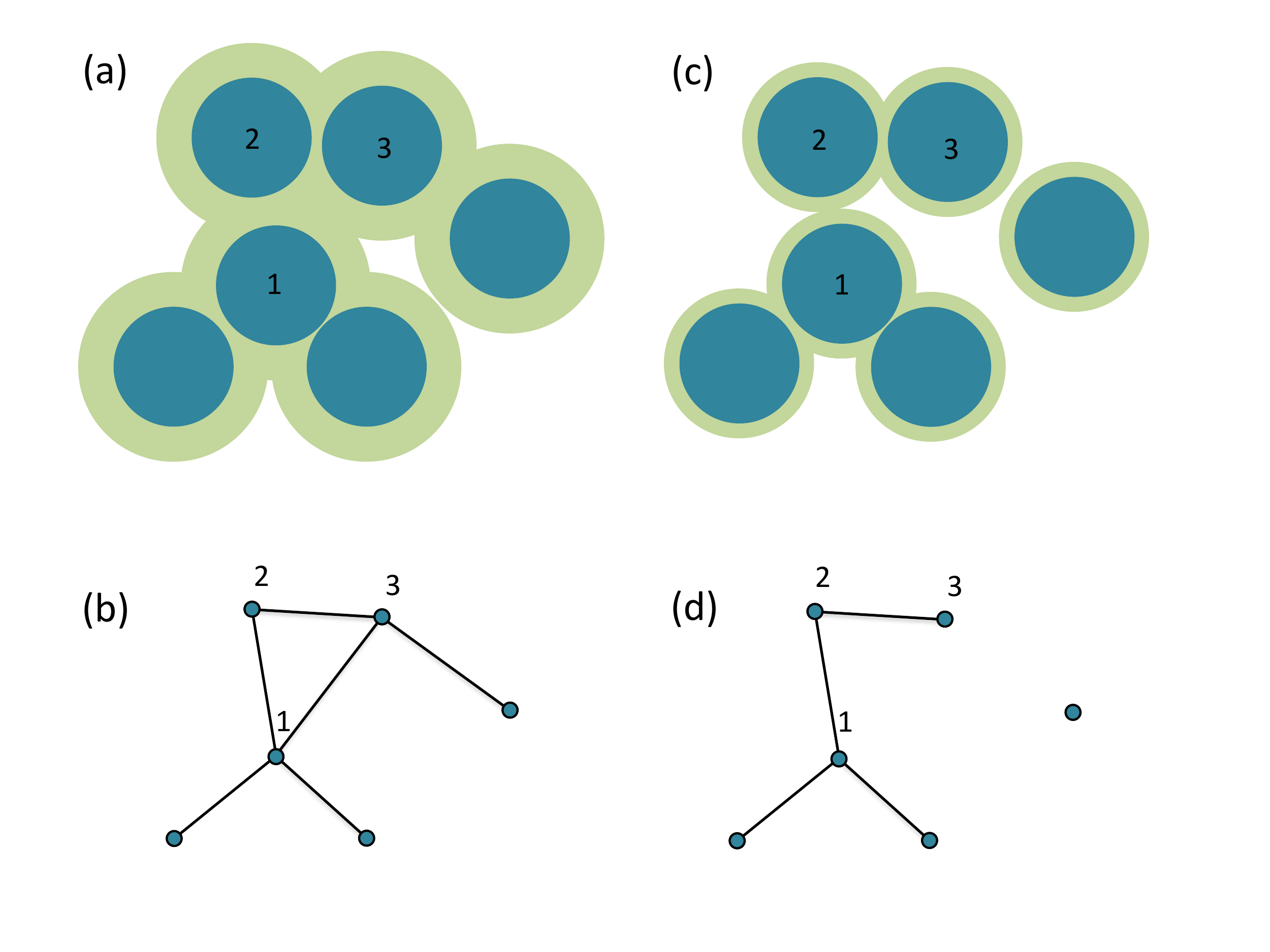}
\caption{Two-dimensional schematic representation of hard spheres with penetrable shells and the corresponding network. In (a) and (c), the 
spherical hard cores are represented by dark circles and the penetrable shells by light annuli surrounding the circles. In (b) and (d) the 
points represent the sphere centers and the edges connecting two points (nodes) are assigned when the corresponding spheres have 
overlapping shells. In (a) the shell thickness is such that the corresponding network in (b) has the nodes labeled by $1$, $2$, and $3$
connected in a closed loop. In (c) the shell thickness is sufficiently small that the corresponding network in (d) has no closed loops. 
}\label{fig1}
\end{center}
\end{figure}

For a given volume fraction of the spheres, the structure of the network so constructed depends crucially on the connectivity range 
$\Delta$ as compared to the hard-core diameter $D$. To see this, let us consider the illustration of Fig.~\ref{fig1}(a), where a set of six hard spheres with their concentric
penetrable shells forms the graph shown in Fig.~\ref{fig1}(b).  The edges between nodes 1, 2, and 3 form a closed loop (a
triangle in this case). The same configuration of the sphere centers with a smaller penetrable shell, Fig~\ref{fig1}(c), gives
rise to the graph of Fig.~\ref{fig1}(d), in which the edge between nodes 1 and 3 is now missing, and the set of node 1, 2, and 3 
no longer form a triangle. Overall, the 6-node graph has now a treelike structure, where closed loops of connected nodes are absent. 
The example of Fig.~\ref{fig1} illustrates the rather intuitive effect that the connectivity range has on the general structure of the network.
Namely, the probability of finding closed loops of connected hard spheres decreases as the connectivity range diminishes.

This effect can be analyzed on more quantitative terms by considering the clustering coefficient $C_3$ (also denoted as the
$3$-node cycle), which is defined as the conditional  probability that two nodes are connected given that they are both connected 
to a third node:\cite{Dall2002,Kong2007} 
\begin{equation}
\label{C3a}
C_3=\frac{\left\langle\sum_{i,j,k}'\theta(\Delta-r_{ij})\theta(\Delta-r_{ik})\theta(\Delta-r_{jk})\right\rangle}
{\left\langle\sum_{i,j,k}'\theta(\Delta-r_{ij})\theta(\Delta-r_{ik})\right\rangle}.
\end{equation}
By introducing the three-particle (or triplet) distribution function,\cite{Hansen2006}
\begin{equation}
\rho^3g^{(3)}(\mathbf{r}_1,\mathbf{r}_2,\mathbf{r}_3)=
\left\langle\sum_{i,j,k}'\delta(\mathbf{r}_1-\mathbf{r}_i)\delta(\mathbf{r}_2-\mathbf{r}_j)\delta(\mathbf{r}_3-\mathbf{r}_k)\right\rangle,
\end{equation}
and by assuming that the system is isotropic and translationally invariant, the cluster coefficient can equivalently be written as
\begin{align}
\label{C3b}
C_3=&\int\! d\mathbf{r} \int\! d\mathbf{r}'g^{(3)}(r,r',\vert \mathbf{r}-\mathbf{r}'\vert)\nonumber \\
&\times
\frac{\theta(\Delta-r)\theta(\Delta-r')\theta(\Delta-\vert\mathbf{r}-\mathbf{r}'\vert))}
{\int d\mathbf{r} \int d\mathbf{r}'g^{(3)}(r,r',\vert \mathbf{r}-\mathbf{r}'\vert)
\theta(\Delta-r)\theta(\Delta-r')}.
\end{align}

In the limit of vanishing hard-core size ($D/\Delta=0$) the system reduces to that of penetrable spheres of diameter $\Delta=\delta$,
where the positions of the sphere centers (or nodes) are completely uncorrelated. In this case,
$g^{(3)}=1$ and by Fourier transforming the Heaviside step functions in \eqref{C3b} we find $C_3=15/32=0.46875$.\cite{Kong2007}
This means that a \textcolor{red}{sizable} fraction of triangles are present in the network, and that this fraction is independent of $\rho$.
The same exercise can be done for closed loops of $n>3$ connected (penetrable) spheres. The corresponding $n$-node
cycles $C_n$, with $n=4$, $5$, $\ldots$, are also independent of the node density and have sizable values. For example, 
$C_4=34/105\simeq0.3238$, $C_5=40949/172032\simeq0.2380$, $C_6=92377/500500\simeq 0.1846$.

Let us now consider the limit $\delta/D\ll 1$ (or equivalently $\Delta/D\simeq 1$). We expand the Heaviside step 
functions in \eqref{C3b} in powers of $\delta$ using $\theta(D+\delta-r)=\theta(D-r)+\delta_d(D-r)\delta+\mathcal{O}(\delta^2)$, where
$\delta_d$ is the Dirac-delta function. Since $g^{(3)}(r,r',\vert \mathbf{r}-\mathbf{r}'\vert)=0$
when $r,r',\vert \mathbf{r}-\mathbf{r}'\vert<D$, we find at the lowest order in $\delta/D$:
\begin{equation}
\label{C3c}
C_3=\frac{\delta}{D}\frac{g^{(3)}(D,D,D)}{\displaystyle\int_{\pi/3}^\pi\!d\theta \sin\theta g^{(3)}(D,D,D\sqrt{2-2\cos\theta})},
\end{equation}
where $\theta$ is the angle between the directions of $\mathbf{r}$ and $\mathbf{r}'$. In the denominator of Eq.~\eqref{C3c}, $g^{(3)}$
is the triplet distribution function for spheres in the rolling contact configuration, in which two particles are allowed to glide
in direct contact over the surface of the third sphere.

\begin{figure}[t]
\begin{center}
\includegraphics[scale=0.58,clip=true]{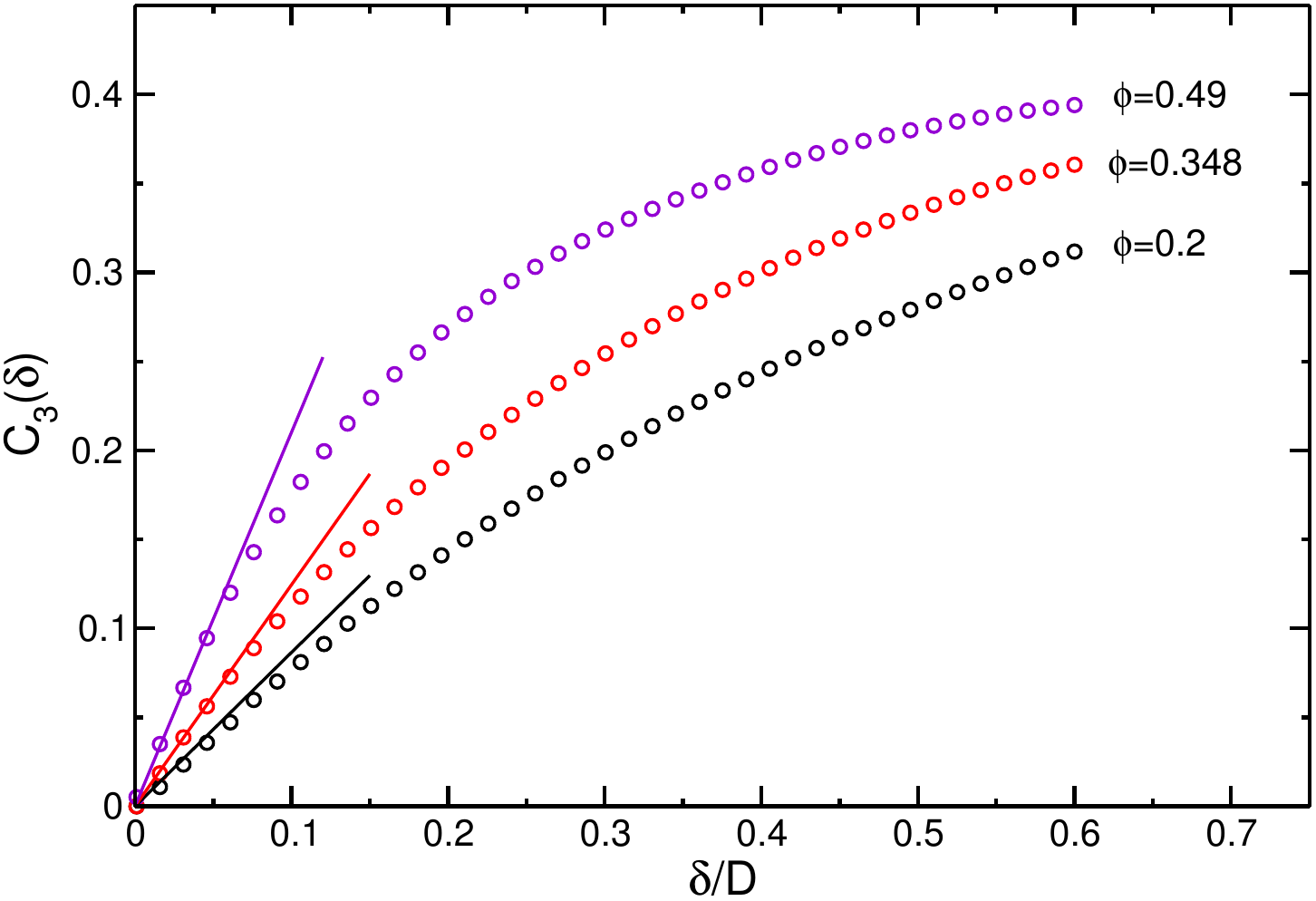}
\caption{Clustering coefficient $C_3$ as a function of the shell thickness for three different values of the volume fraction $\phi$.
Symbols are Monte Carlo calculations of Eq.~\eqref{C3a}, while the solid lines are obtained from Eq.~\eqref{C3d}.}\label{fig2}
\end{center}
\end{figure}

For suspensions of hard spheres at equilibrium, the value of the three-particle distribution function at contact, $g^{(3)}(D,D,D)$, increases for larger
concentrations of spheres. It remains however finite as long as the volume fraction is smaller than the liquid-solid state transition point at $\phi^*\simeq 0.494$.
For $\phi<\phi^*$ therefore the clustering coefficient becomes vanishingly small as $\delta/D\rightarrow 0$. To see in details the effect of $\phi$ on
$C_3$, we adopt the Attard's expression for the triplet distribution function of hard-spheres in the rolling contact configuration:\cite{Attard1991}
\begin{equation}
\label{attard}
g^{(3)}(D,D,D\sqrt{2-2\cos\theta})\simeq g^{(2)}(D)^2\frac{g^{(2)}(s(\theta))+1}{2},
\end{equation}
where $g^{(2)}$ is the pair distribution function and $s(\theta)=D(1+\theta-\pi/3)$. By substituting Eq.~\eqref{attard}
in Eq.~\eqref{C3c}, and by denoting $I=\int_{\pi/3}^\pi\!d\theta\sin\theta [g^{(2)}(s(\theta))+1]/2$, we obtain: 
\begin{equation}
\label{C3d}
C_3\simeq \frac{\delta}{D}\frac{g^{(2)}(D)+1}{2I}\simeq \frac{5}{16}\frac{\delta}{D}\left[\frac{1-\phi/2}{(1-\phi)^3}+1\right],
\end{equation}
where in the last expression we have used the Carnahan-Starling approximation for $g^{(2)}(D)$ and we have set 
$I\simeq 8/5$ (see Appendix \ref{appa}). Despite its simplicity, 
Eq.~\eqref{C3d} reproduces well the numerical results of the clustering coefficient in the $\delta/D\ll 1$ regime.
This is shown in Fig.~\ref{fig2}, where Eq.\eqref{C3d} (solid lines) is compared to the results of Monte Carlo calculations of 
Eq.~\eqref{C3a} (open symbols). In performing the configurational averages, we have used the Metropolis algorithm to generate  
$300$ equilibrated configurations of $N=2000$ hard-spheres for each value of $\phi$. 

Following the same steps that lead to Eq.~\eqref{C3c}, it is easy to show that for loops formed by $n$ particles, the
corresponding $n$-cycle coefficient scales as $C_n\propto \delta/D$ for $\delta/D\ll 1$, indicating that closed loops 
of any $n$ become negligible in this limit.

\section{Tree-ansatz percolation}
\label{percolation}

The result of the Sec.\ref{clustering} implies that for sufficiently thin penetrable shells the network
formed by connected hard spheres has essentially a treelike structure. This property can be exploited to calculate
the percolation threshold of equilibrium distributions of hard spheres under the assumption that $\delta/D\ll 1$. 

We start by defining the degree distribution of a node (or sphere) as the probability $P(k)$ that a randomly selected node is connected
to exactly $k$ other nodes of the network. Let us assign the location of the randomly selected node at $\mathbf{r}_1$. Since 
a node located at, say, $\mathbf{r}_j$ is connected (not connected) to the 
selected node if the distance $r_{1j}=\vert \mathbf{r}_1-\mathbf{r}_j\vert$ is smaller (larger) than $\Delta$, the degree
distribution $P(k)$ takes the following form:\cite{Bradonjic2008}
\begin{equation}
\label{degree1}
P(k)=\binom{N-1}{k}\left\langle\prod_{i=2}^{k+1}\theta(\Delta-r_{1i})\prod_{j=k+2}^N\theta(r_{1j}-\Delta)\right\rangle,
\end{equation}
where the configurational average makes $P(k)$ independent of the node labels and the binomial factor gives the number of
ways that $k$ unordered nodes are chosen from a total of $N-1$ nodes. 

\begin{figure}[b]
\begin{center}
\includegraphics[scale=0.58,clip=true]{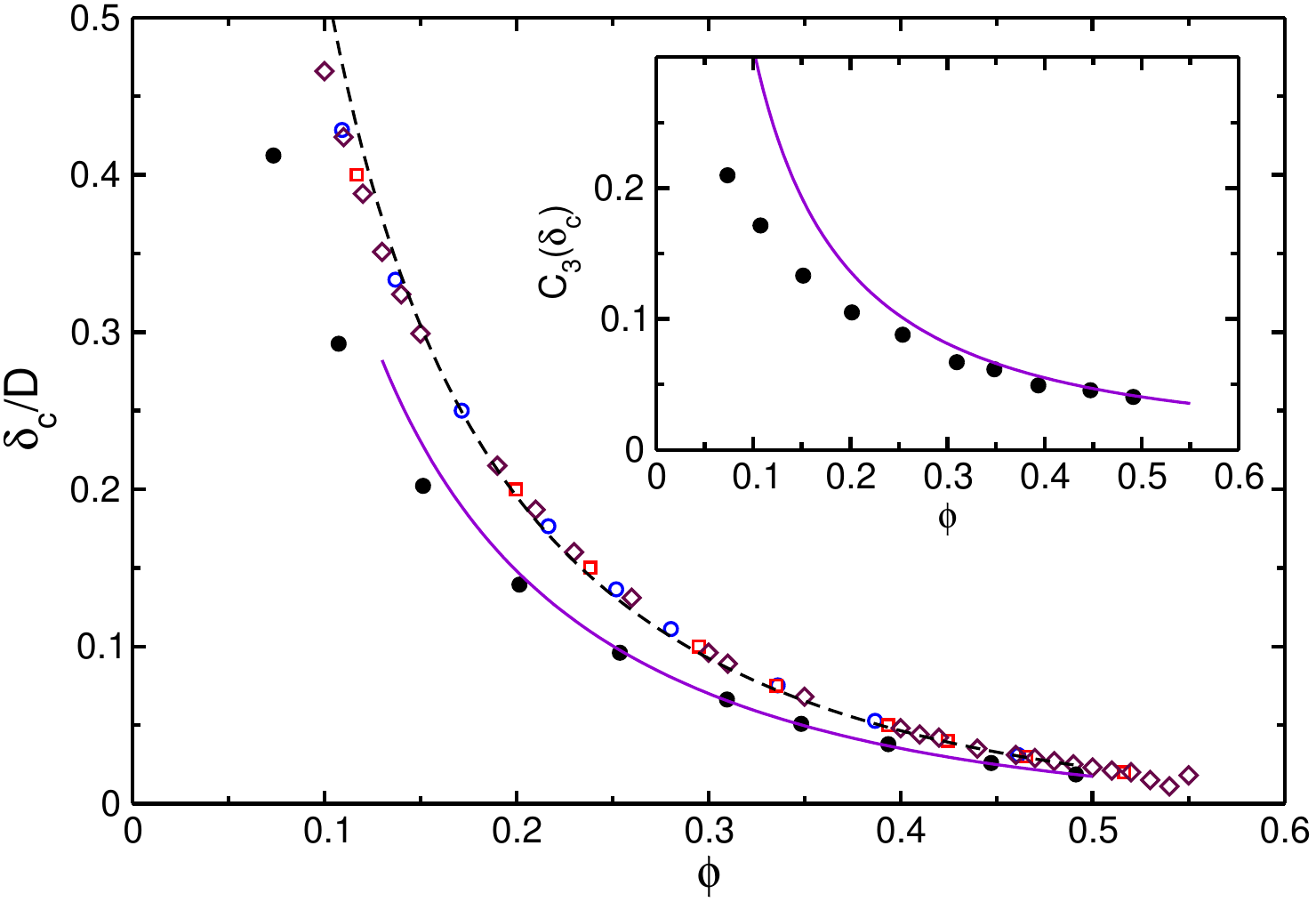}
\caption{Critical distance $\delta_c$ as a function of the volume fraction $\phi$ of the hard spheres. Filled circles are the 
tree-ansatz results obtained from the numerical calculation of Eq.~\eqref{pt1}. The analytical formula \eqref{pt4} is plotted by 
the solid line. The dashed line is Eq.~\eqref{pt4} plotted with a different prefactor (see text).
Monte Carlo results for $\delta_c$ are shown by open diamonds,\cite{Heyes2006} open squares,\cite{Miller2009} and 
open circles.\cite{Ambrosetti2010a} Note that the Monte Carlo results for $\phi>0.49$ are in the metastable region. Inset: solid circles
are the numerical results of the clustering coefficient calculated at the tree-ansatz percolation threshold. The solid line is
$C_3=[1+(1-\phi)^3/(1-\phi/2)]/54\phi$, which is obtained from Eqs.~\eqref{C3d} and \eqref{pt4}. 
}\label{fig3}
\end{center}
\end{figure}

Knowledge of $P(k)$ allows us to calculate the mean size $S$ of a cluster (or component) to which a randomly selected node 
belongs. For infinite systems, and if there is no giant component in the graph, the divergence of $S$ marks the percolation 
transition, that is, the transition at which a giant cluster of connected nodes first appears.\cite{Newman2001} 
Under the assumption that the network is treelike, the component to which the
selected node belongs is formed by branches attached to the node according to the degree distribution $P(k)$. Hence:
\begin{equation}
\label{S1}
S=1+\sum_k P(k)k T=1+\langle k\rangle T,
\end{equation}
where $\langle  k\rangle =\sum_k kP(k)$ and
$T$ is the mean size of one of the $k$ branches attached to the selected node. Owing to the treelike
structure of the network, $T$ is given by the mass (unity) of one neighbor of the selected node plus the mean size of each
of the remaining $k-1$ subbranches attached to the neighbor node:
\begin{equation}
\label{T1}
T=1+\sum_k Q(k)(k-1)T,
\end{equation}
where $Q(k)$ is the degree distribution of a node that is a neighbor of the selected node. To find $Q(k)$ we note that if
we select at random an edge directly connecting two nodes, and we follow the edge from one node to its neighbor, 
the node that we arrive at by following that edge will be $k$ times more likely to have degree $k$ than degree $1$. 
Its degree distribution will thus be proportional to $kP(k)$, which implies that, after suitable normalization, 
$Q(k)=kP(k)/\langle k\rangle$. Substituting this
expression in Eq.~\eqref{T1}, and defining $\langle  k^2\rangle =\sum_k k^2P(k)$, we arrive at the condition for the divergence of
$S$:\cite{Newman2001,Boccaletti2006}
\begin{equation}
\label{pt1}
\frac{\langle  k^2\rangle}{\langle  k\rangle}=2.
\end{equation}
From the degree distribution of Eq.~\eqref{degree1}, we express $\langle  k\rangle$ and $\langle  k^2\rangle$ in terms
of configurational averages of the hard-sphere system as shown in Appendix \ref{appb}. In this way, the condition \eqref{pt1} reduces 
to:
\begin{equation}
\label{pt2}
\frac{\left\langle\sum_{i,j,k}'\theta(\Delta-r_{ij})\theta(\Delta-r_{ik})\right\rangle}{\left\langle\sum_{i,j}'\theta(\Delta-r_{ij})\right\rangle}=1,
\end{equation}
which identifies the percolation threshold in terms of the critical density $\rho_c$ for a given $\Delta$ or, equivalently, the critical
distance $\Delta_c=\delta_c+D$ for a given $\rho$. We adopt the latter definition to calculate numerically $\delta_c/D$ as a function
of the volume fraction $\phi$. The configuration averages in Eq.~\eqref{pt2} are performed as described in Sec.~\ref{clustering} by choosing 
for a given $\phi$ an initial value $\delta_\textrm{init}$ within the interval $\Delta\delta=\delta_\textrm{max}-\delta_\textrm{min}$, 
where $\delta_\textrm{min}=0$ and $\delta_\textrm{max}$ is large enough so that the left-hand side of Eq.~\eqref{pt2} is surely larger than $1$. 
The critical distance $\delta_c$ is then found by bisecting the interval $\Delta\delta$ until the condition \eqref{pt2} is satisfied. 
The resulting values of $\delta_c/D$ are plotted in Fig.~\ref{fig3} by filled circles and are compared with previously published
Monte Carlo calculations of the percolation threshold reported by open symbols.\cite{Heyes2006,Miller2009,Ambrosetti2010a} 
Figure \ref{fig3} confirms our conjecture that the tree-ansatz approach is increasingly accurate as $\delta_c/D$ decreases. 
Note, however, that as the system approaches the liquid-solid phase transition, there is a residual discrepancy  between the tree-ansatz
percolation and the exact numerical results. This is due to the small but non-zero clustering coefficient at percolation, shown 
in the inset of Fig.~\ref{fig3}, whose value is about $0.04$ for the density closest to the liquid-solid transition. 

To find an analytic expression of $\delta_c/D$ within the tree-ansatz approximation, we express Eq.~\eqref{pt2} in terms of the three-
and two-particle distribution functions. At the lowest order in $\delta_c/D$, the left-hand side of
Eq.~\eqref{pt2} reduces to:
\begin{align}
\label{pt3}
&\frac{\displaystyle\rho\int\!d\mathbf{r}\int\!d\mathbf{r}'g^{(3)}(r,r',\vert \mathbf{r}-\mathbf{r}'\vert)\theta(\Delta_c-r)\theta(\Delta_c-r')}
{\displaystyle\int\!d\mathbf{r} g^{(2)}(r)\theta(\Delta_c-r)}\nonumber \\
&\simeq\frac{\displaystyle 2\pi\rho\delta_c D^2\int_{\pi/3}^\pi\!d\theta\sin\theta\,g^{(3)}(D,D,D\sqrt{2-2\cos\theta})}{g^{(2)}(D)}.
\end{align}
We express $g^{(3)}$ as in Eq.~\eqref{attard} and approximate the remaining integration over $\theta$ by $I\simeq 8/5$, as done in Sec.~\ref{clustering}.
We next equate the result to $1$ to find the following expression for the critical distance:
\begin{equation}
\label{pt4}
\frac{\delta_c}{D}\simeq \frac{1}{12 I\phi g^{(2)}(D)}\simeq\frac{5}{96}\frac{(1-\phi)^3}{\phi (1-\phi/2)},
\end{equation}
where we have again used the Carnhan-Starling approximation for $g^{(2)}(D)$. Equation \eqref{pt4} (solid line in Fig.~\ref{fig3}) converges
asymptotically to the numerical solution of Eq.~\eqref{pt2} as $\phi$ increases towards the liquid-solid transition point, and it provides
therefore an approximate analytical expression for the critical distance in that density region. By replacing in Eq.~\eqref{pt4}
the numerical prefactor $5/96\simeq 0.052$ by $1.65/24\simeq 0.06875$, we recover the expression that was used 
in Refs.~\onlinecite{Ambrosetti2010a,Grimaldi2015b} to approximate the critical distance for a much wider region of the hard sphere 
volume fraction (dashed line in Fig.~\ref{fig3}). Equation \eqref{pt4} provides therefore a new interpretation of the  
approximation scheme adopted in Refs.~\onlinecite{Ambrosetti2010a,Grimaldi2015b}, which derived from the observation that $\delta_c$ 
is approximately proportional to the mean separation between nearest-neighbors hard spheres for $\phi\gtrsim 0.1$.

\section{conclusions}
\label{concl}

Following the observation that the probability of finding closed loops in the network of connected hard spheres 
becomes smaller as the connectivity distance decreases, we have calculated the percolation threshold in the limit in which
the percolating cluster has a tree-like structure. Our results agree well with Monte Carlo calculations of the critical threshold when the
connectivity distance is sufficiently smaller than the hard core diameter. The tree-ansatz formalism presented here differs
from other approximation schemes in that it takes into full account, in the pair and triplet distribution functions, the structural 
correlations of the percolating particles.
We conclude by pointing out that the tree-ansatz approach could be extended to describe percolation also in systems of anisotropic 
particles like, e.g., rod and platelet particles. Suitable expressions for the pair and triplet distribution functions
of anisotropic particles are however required to obtain explicit formulas for the percolation threshold.


\appendix
\section{Calculation of $I$}
\label{appa}

\begin{figure}[t]
\begin{center}
\includegraphics[scale=0.58,clip=true]{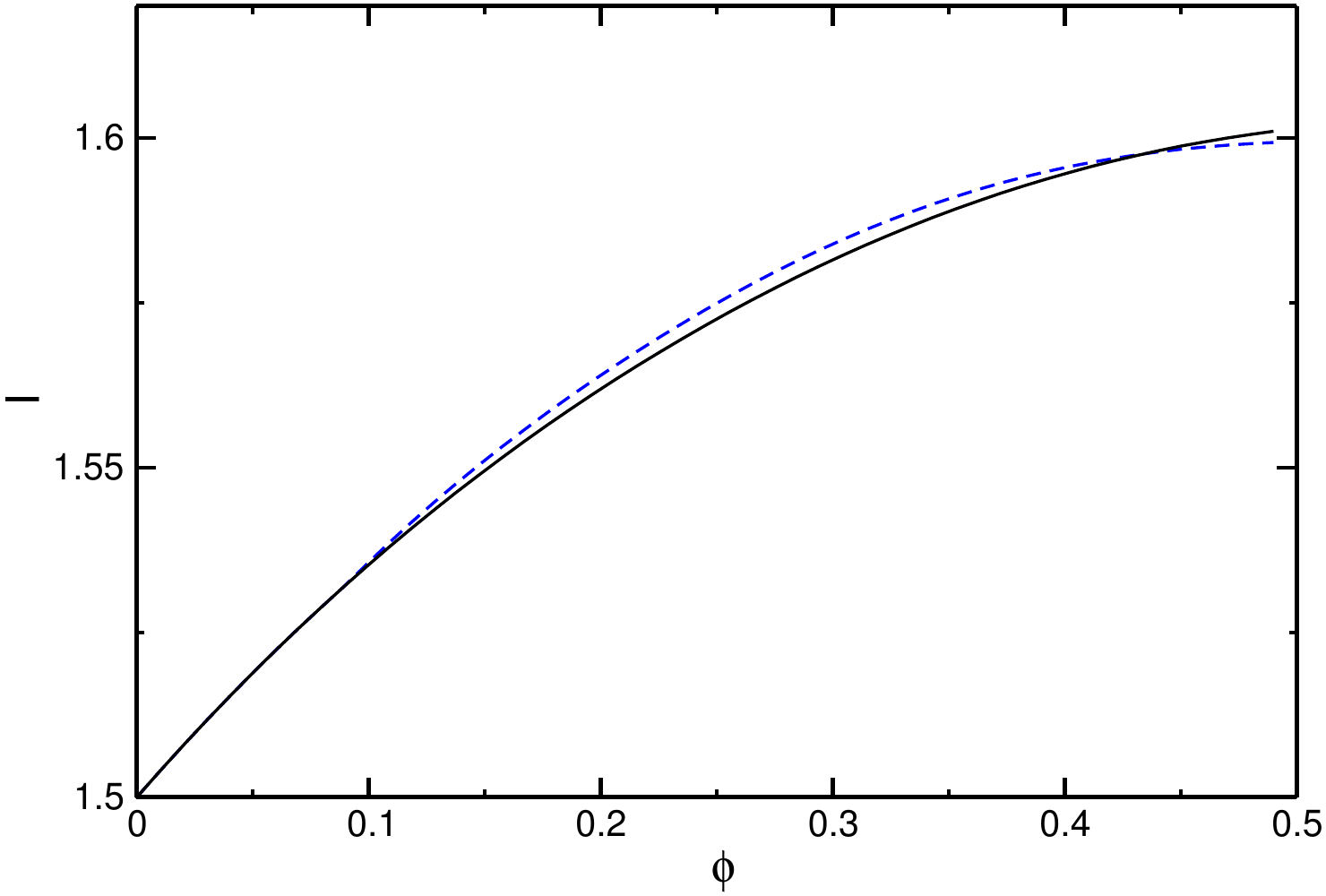}
\caption{The function $I$, Eq.~\eqref{appa1}, obtained by numerical integration using two different approximations for 
the pair distribution function.}\label{fig4}
\end{center}
\end{figure}

Using the approximation given in Eq.~\eqref{attard} for the three-particle distribution function in the rolling contact 
configuration, the clustering coefficient and the percolation threshold within the tree-ansatz depend on the following
integral:
\begin{equation}
\label{appa1}
I=\int_{\pi/3}^\pi\!d\theta \sin\theta \frac{1+g^{(2)}(s(\theta))}{2},
\end{equation}
where $g^{(2)}$ is the pair distribution function and $s(\theta)=D(1+\theta-\pi/3)$. Figure \ref{fig4} shows numerical
calculations of $I$ with $g^{(2)}$ given by the Percus-Yevick approximation \cite{Hansen2006, PY} (solid line) and by the empirical formula of
Ref.~\onlinecite{Trokhymchuk2006} (dashed line). The integral $I$ depends very weakly on the volume fraction $\phi$, and it
ranges from $I=3/2$ at $\phi=0$ to $I\simeq 1.6$ at $\phi=0.49$. For values of $\phi$ close to the liquid-solid phase transition, we
adopt $I=1.6=8/5$.

\section{Evaluation of $\langle k\rangle$ and $\langle k^2\rangle$}
\label{appb}

We follow Ref.~\onlinecite{Newman2001} to calculate the first and second moments of 
the degree distribution $P(k)$ given in Eq.~\eqref{degree1}. To this end, we first introduce the
generating function of $P(k)$ defined as: 
\begin{equation}
\label{appb1}
G(x)=\sum_k P(k)x^k,
\end{equation}
from which $\langle k\rangle$ and $\langle k^2\rangle$ are given by
\begin{align}
\label{appb2a}
&\langle k\rangle=\sum_k k P(k)=\left[\frac{d G(x)}{dx}\right]_{x=1}, \\
\label{appb2b}
&\langle k^2\rangle=\sum_k k^2 P(k)=\left[\frac{d G(x)}{dx}\right]_{x=1}+\left[\frac{d^2 G(x)}{dx^2}\right]_{x=1}.
\end{align}
Using Eqs.~\eqref{degree1} and \eqref{appb1}, the generating function can be expressed as:
\begin{align}
\label{appb3}
G(x)&=\left\langle\prod_{j=2}^{N-1}[x\theta(\Delta-r_{1j})+\theta(r_{1j}-\Delta)]\right\rangle \nonumber \\
&=\frac{1}{N}\left\langle\sum_i\prod_{j\neq i}[x\theta(\Delta-r_{ij})+\theta(r_{ij}-\Delta)]\right\rangle.
\end{align} 
Applying Eqs.~\eqref{appb2a} and \eqref{appb2b} to 
Eq.~\eqref{appb3} leads to:
\begin{align}
&\langle k\rangle=\frac{1}{N}\left\langle\sum_{i,j}'\theta(\Delta-r_{ij})\right\rangle, \\
&\langle k^2\rangle=\langle k\rangle+\frac{1}{N}\left\langle\sum_{i,j,k}'\theta(\Delta-r_{ij})\theta(\Delta-r_{ik})\right\rangle,
\end{align}
which when substituted in Eq.~\eqref{pt1} give Eq.~\eqref{pt2}.

\end{document}